\documentclass[12pt]{iopart}
\usepackage{iopams}
\usepackage{graphicx}% Include figure files

\begin{document}

\title[Center-of-mass  motion in TDHF]{Center-of-mass motion and cross-channel coupling in time-dependent Hartree-Fock theory}

\author{A S Umar and V E Oberacker}
\address{Department of Physics and Astronomy, Vanderbilt University,
         Nashville, Tennessee 37235, USA}
\ead{umar@compsci.cas.vanderbilt.edu}

\begin{abstract}
We provide a discussion of issues related to the center-of-mass motion and cross-channel coupling
in applications of the time-dependent Hartree-Fock (TDHF) theory to heavy-ion collisions. We find
that the entrance channel dynamics of a heavy-ion collision as described by TDHF does not seem to
be significantly influenced by these effects, whereas the long-time evolution may be less reliable.
\end{abstract}
\submitto{\JPG}
%\maketitle

%------------------------------------------------------------------------------

\section{Introduction}
It is generally acknowledged that the time-dependent Hartree-Fock (TDHF) theory provides a
useful foundation for a fully microscopic many-body theory of low-energy heavy-ion reactions~\cite{Ne82}.
The success of the TDHF method is predicated on the expectation that the Pauli principle plays an important role in
simultaneously building up a time-dependent mean-field and suppressing the propagation of the
strong $N-N$ interaction terms.
In recent years there has been a significant progress in the study of heavy-ion collisions
using the TDHF theory~\cite{UO06a,MR06a,SC01}. This progress is partially due to the available computational
power which allows calculations without resorting to simplifying but
somewhat unphysical assumptions used in the past, as well as to the considerable improvements
made in the parametrization of effective interactions~\cite{CB98}.

One of the issues faced in
the reduction of the finite many-body problem to be described in terms of single-particle
degrees of freedom and a mean-field is the center-of-mass (c.m.) motion.
This is due to the fact that the resulting wavefunction, say a Slater determinant,
does not factorize into a product of one single center-of-mass wave-function and a wavefunction of the
internal coordinates~\cite{DF74}.
Thus the calculated energies include a contribution
from the center-of-mass motion in a complicated way. Another way of describing the problem
is that the coordinates of the nucleons in the intrinsic wavefunction, $\chi(\mathbf{r}_1,\mathbf{r}_2,\ldots,\mathbf{r}_A)$,
should be defined with respect to a fixed origin, which means the coordinates are not independent
and satisfy $\sum _{i=1}^{A}\mathbf{r}_{i}=0$.

The second problem faced in TDHF has to do with exit channel properties for collisions where more than one final
fragment exists, e.g. deep-inelastic collisions. In this case, although we may have two well separated fragments,
the final Slater determinant cannot be transformed to a block diagonal form thus making the identification of asymptotic
channels unclear. This is known as the {\it cross-channel coupling} problem.

In the sections below we shall discuss the ramifications of these issues to the interpretation of
the results obtained from TDHF calculations. In particular we shall discuss each problem in a more
detailed fashion and show the magnitude of the effect and try to establish the conditions under
which these issues may have a minimal effect on the calculated observables.

\section{C.M. Motion and TDHF}
For many-body wavefunctions obtained using variational calculations
there is no well defined prescription for eliminating the center-of-mass coordinate.
Most of the suggested methods can be described by the general expression given by Lipkin~\cite{Li58}
\begin{equation}
\Psi(\mathbf{r}'_1,\ldots ,\mathbf{r}'_{A-1})=\int\; g(\mathbf{R})\Phi(\mathbf{r}_1,\ldots,\mathbf{r}_A)d\mathbf{R}\;,
\label{eq:wave}
\end{equation}
where the function $\Phi$ is the general wavefunction depending on the coordinates of $A$ nucleons,
the function $g(\mathbf{R})$ is the weight function for integrating over the center-of-mass coordinate $\mathbf{R}$,
and the result is the wavefunction $\Psi$ depending on the $3(A-1)$ internal coordinates.
An equivalent expression in momentum space is given in Ref.~\cite{PT62}.
One common derivative of the above is to project out the $\mathbf{P}=0$ part of the many-body
wavefunction.
The proper way of choosing the weight function, $g(\mathbf{R})$, is such that
the expectation value of the internal Hamiltonian (i.e. the Hamiltonian after
the removal of the center-of-mass dependence) for the wavefunction (\ref{eq:wave})
is a minimum. A proof for the above statement is given in Section 4 of Ref.~\cite{Bouten}.
In practice this turns out to be a very difficult task. Alternatively, one can
specify the functional form for $g(\mathbf{R})$ but include
variational parameters, similar to the Lagrange parameters for constraints,
in the definition. The variational minimization then would also include these parameters
thus giving the best $g(\mathbf{R})$ for the chosen functional form~\cite{Bouten}.
While these methods may be useful for the calculation of static properties, a generalization
to dynamical calculations poses serious technical problems and therefore are not used in
practice. As discussed below, TDHF initial wavefunctions actually need the center-of-mass
wavefunction to allow for boosts.

If one is primarily interested in energies an alternate procedure is the subtraction of the
center-of-mass energy from the total kinetic energy of the system. If this is done during the
minimization procedure one may effectively achieve the above task.
Formally, this can be written as:
\begin{eqnarray*}
\hat{K}-\hat{K}_{c.m.}&=&\sum_{i=1}^{A}\frac{\mathbf{p}_i^2}{2m}-\frac{\left(\sum_{i=1}^{A}\mathbf{p}_i\right)^2}{2mA}\\
\ &=&\frac{1}{2m}\left[1-\frac{1}{A}\right]\sum_{i=1}^{A}\mathbf{p}_i^2-\frac{1}{2mA}\sum_{i \neq j}\mathbf{p}_i\cdot\mathbf{p}_j\;.
\end{eqnarray*}
The last part of this expression is numerically difficult as it includes double integrals and is
usually neglected. Consequently, most Skyrme forces use the simple correction to the kinetic energy
given in the first term. There are various Skyrme fits that include the quadratic term as well,
such as the SLy6 parametrization~\cite{CB98}. The general statements made about this correction is that it
is not very sensitive to deformation but it can be substantial in magnitude. A comprehensive study of the influence
of the center-of-mass correction in mean-field studies is given in Ref.~\cite{BR00}. This problem
has been recently raised in Ref.~\cite{JS07} in the context of calculating fission barriers,
where a corrective procedure was suggested.

In TDHF, the static Hartree-Fock Slater determinant is boosted by multiplying each Slater
determinant with an exponential phase factor
\begin{equation}
\Phi _{j}\rightarrow \exp (i\mathbf{k}_{j}\cdot \mathbf{R}_j)\Phi _{j}\;,
\label{eq:boost}
\end{equation}
where $\Phi _{j}$ is the HF wavefunction for nucleus $j$, $\mathbf{R}_j$ is the corresponding
center of mass coordinate
\begin{equation}
\mathbf{R}_j=\frac{1}{A_{j}}\sum _{i=1}^{A_{j}}\mathbf{r}_{i}\;,
\end{equation}
and $\mathbf{k}_{j}$ is the boost momentum chosen to be tangent to the Coulomb trajectory
at an initial separation distance.
The target and projectile are placed to be in the center-of-mass frame for the entire
system and satisfy
\begin{equation}
A_T\mathbf{R}_T+A_P\mathbf{R}_P=0\;.
\end{equation}
Thus, the total momentum for the system has expectation value zero, and the relative
momentum is correctly given by $\mathbf{k}_T-\mathbf{k}_P$. Here, the assumption is that the
physics of the collision is determined by the {\it relative momenta}. This is one
of the reasons why we expect the contributions coming from the wave packets contained
in the initial Slater determinants to a large extent cancel out. However, as we discuss
below, there are other reasons as to why projecting out the center-of-mass part of the
wavefunctions is not appropriate in TDHF calculations.
\begin{figure}[!htb]
\begin{center}
\includegraphics*[scale=0.42]{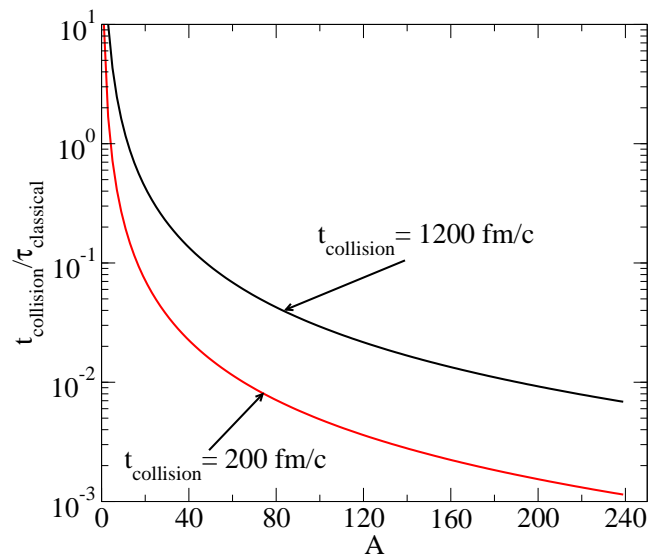}
\caption{\label{fig:wp} The ratio $t_{\textrm{collision}}/\tau_{\textrm{classical}}$ plotted
as a function of mass number $A$, for two collision times.}
\end{center}
\end{figure}

In TDHF the initial boosted states are interpreted as wave packets because the wave packet is already
contained in the static Slater determinants~\cite{FKW,Ir80}. The result of multiplication by the phase
factor given above is to shift the center of momentum of the wave packet by the vector {\bf k}.
However, there are fundamental differences between a boosted TDHF solution and a Schr\"odinger wave packet.
Since we are using a single Slater determinant we are not
free to change the wave packet nature of the solution. The two are coupled via the
minimization process and this goes beyond just the contribution/correction to static
energies. This coupling results in a {\it non-spreading} wave packet for the TDHF evolution~\cite{Ne82}.
It can be shown that the TDHF wave packet is identical to the Schr\"odinger wave packet when
the terms causing the spreading in the Schr\"odinger wave packet are neglected~\cite{US84,FKW}.
The condition that the wave packet behave completely classically over the time scale of the
collision requires~\cite{US84,GW75} (i.e. the spreading of the wave packet never
catches up with its kinematical motion)
\begin{equation}
\frac{t_{\textrm{collision}}}{\tau_{\textrm{classical}}}\ll 1 \;\;\;\;\;\;,\;\;\;\;\; \tau_{\textrm{classical}}=
\left[\frac{\hbar}{M\sigma^2}\right]^{-1}\;,
\label{eq:wp}
\end{equation}
where $\sigma$ is the root-mean-square width of the Gaussian wave packet representing the nucleus,
and $M$ is the total mass of the nucleus.
If we take the initial
wave packet to be characterized by the size of the nucleus we can write
\begin{equation}
\tau_{\textrm{classical}}=\frac{2mc^2r_0^2A^{5/3}}{\hbar c}\approx 19A^{5/3}\; fm/c\;.
\end{equation}
Corresponding spreads in c.m. momentum and energy are given by~\cite{Ne82}
\begin{equation}
\Delta p_{c.m.} > \frac{\hbar}{2r_0A^{1/3}}\;,
\end{equation}
with the corresponding spread in beam-energy
\begin{equation}
\frac{\Delta E}{E}=\frac{\hbar}{(2mEr_0A^{4/3})^{1/2}}\;,
\end{equation}
where $E$ is the beam energy per particle. For light ions this spread could be
as much as 10\% but reduces for heavier systems.

In Fig.~\ref{fig:wp} we plot the dependence of the ratio in Eq.~(\ref{eq:wp}) on the mass number for two
different reaction times. The longer time scale of $1200$ fm/c refers to the whole reaction,
while the shorter time scale of $200$ fm/c describes the initial evolution until the nuclei begin to overlap.
We see that for heavy systems the ratio is small and the
TDHF wave packet may be considered as the classical limit of a Schr\"odinger wave packet.
On the other hand, for long reaction times and light systems the correspondence does not seem to be
very good. However, this argument may be flawed since one can take the position that we are
only interested in the evolution of the wave packet until the two nuclei begin to have substantial overlap~\cite{FKW}.
In this case the collision times are much smaller thus resulting in a smaller ratio for most
mass numbers, as shown in Fig.~\ref{fig:wp}.
In this context one can interpret the TDHF results from a classical perspective, which
is the correct limit for a non-spreading wave packet.
Consequently, center-of-mass corrections are not appropriate in TDHF calculations and have always
been neglected. The effect of using Skyrme forces fitted with the simple c.m. correction term
without this correction in TDHF will be discussed further in the next section.
\begin{figure}[!htb]
\begin{center}
\includegraphics*[scale=0.45]{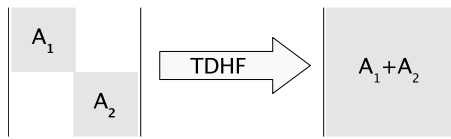}
\caption{\label{fig:tdhf_collision} Schematic illustration of the initial and final many-body states.
The initial state is block diagonal whereas the final state is a full Slater determinant.}
\end{center}
\end{figure}
Implications of the above uncertainties depend on the details of the collision process
under study. Reactions that are predominantly determined by the entrance channel, such as
the initial formation of the compound system in fusion, or the calculation of fusion barriers using
the density-constrained TDHF (DC-TDHF) method~\cite{UO06b} are expected to be most reliable.
On the other hand longer time processes such as deep-inelastic scattering will be prone to
higher uncertainties. For example, in calculating Wilczynski plots (double-differential
cross-section with respect to scattering angle and energy) for deep-inelastic collisions
we may expect a spread in energy, as much as $10$~MeV for light systems and a spread in
impact parameter of order $1$~fm.

\section{Cross-Channel Coupling}
The so-called {\it cross-channel coupling} describes the problem that
in deep-inelastic TDHF collisions the final fragments remain {\it entangled}.
This complicates the identification of exit channel fragments as distinct nuclear
systems~\cite{KD77}.
In TDHF, the entrance channel asymptotic conditions are set up correctly as discussed in the
previous section.
In practice, the initial nuclei are calculated using the static Hartree-Fock theory without
the center-of-mass correction. The resulting Slater
determinants for each nucleus comprise the larger Slater determinant describing the colliding
system during the TDHF evolution, as depicted in Fig.~\ref{fig:tdhf_collision}.
Nuclei are assumed
to move on a pure Coulomb trajectory until they reach the initial separation between the nuclear centers used
in the TDHF evolution. Using the Coulomb trajectory we compute the relative kinetic energy at this
separation and the associated translational momenta for each nucleus in the center-of-mass
frame. The nuclei are then boosted by multiplying each HF determinant by an exponential phase
as in Eq.~(\ref{eq:boost}) so that the relative momentum is the correct one calculated for the
Coulomb collision.
The Galilean invariance of the TDHF equations with the full Skyrme force results in the evolution of
the system without spreading and the conservation of the total energy for the system.
In other words, a translating Slater determinant does not dissipate its kinetic energy.
In TDHF, the many-body state remains a Slater determinant at all times. The final state
is a filled determinant, even in the case of two well separated fragments (does not go back to
block-diagonal form). This phenomenon
is commonly known as the {\it cross-channel coupling} and indicates that it is not possible
to identify the well separated fragments as distinct nuclei since each single particle state
will have components distributed everywhere in the numerical box. In this sense it is
only possible to extract {\it inclusive} (averaged over all states) information from these calculations.

\begin{figure*}[!hbt]
\begin{center}
\includegraphics*[scale=0.35]{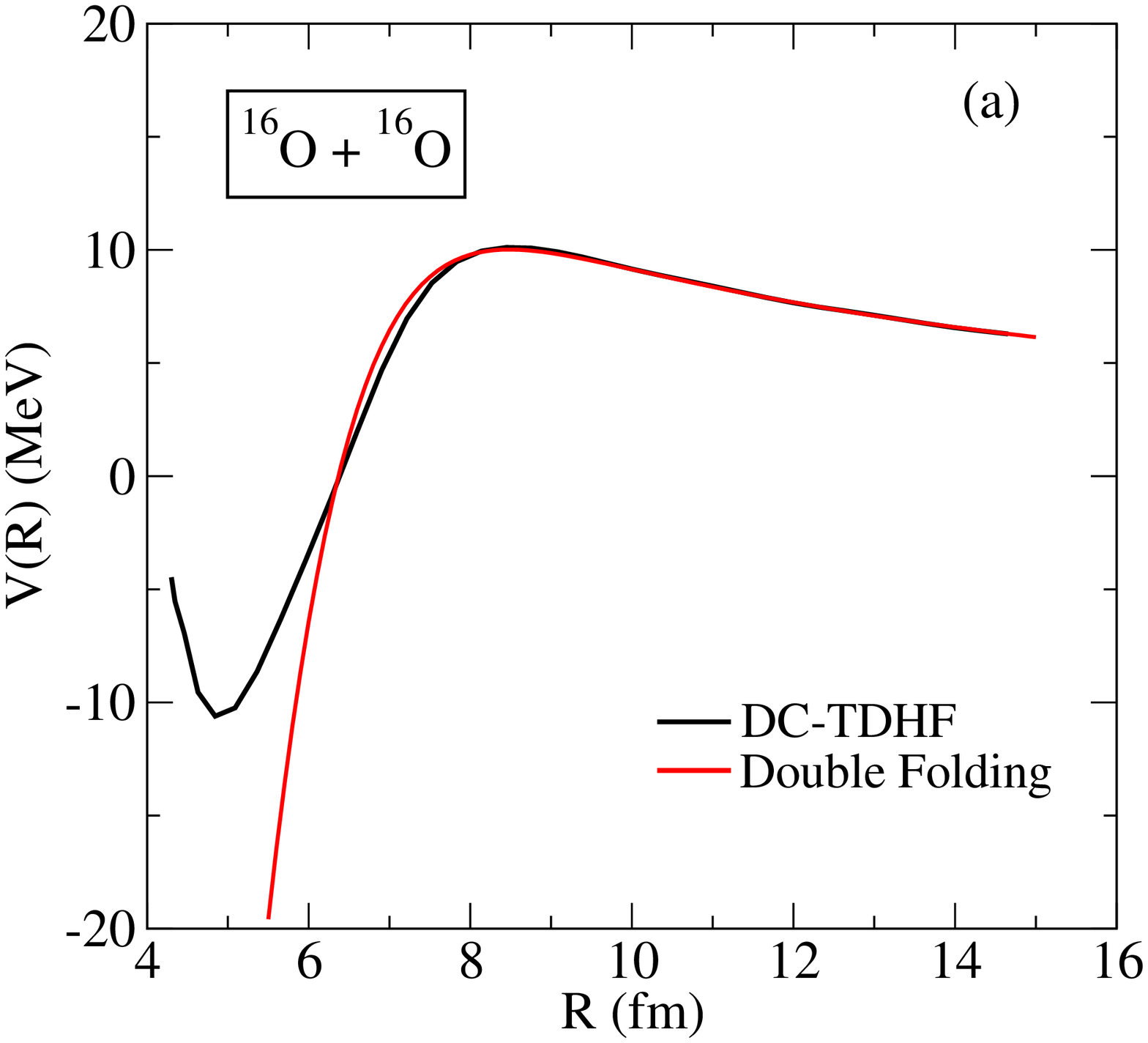}\hspace{0.1in}\includegraphics*[scale=0.35]{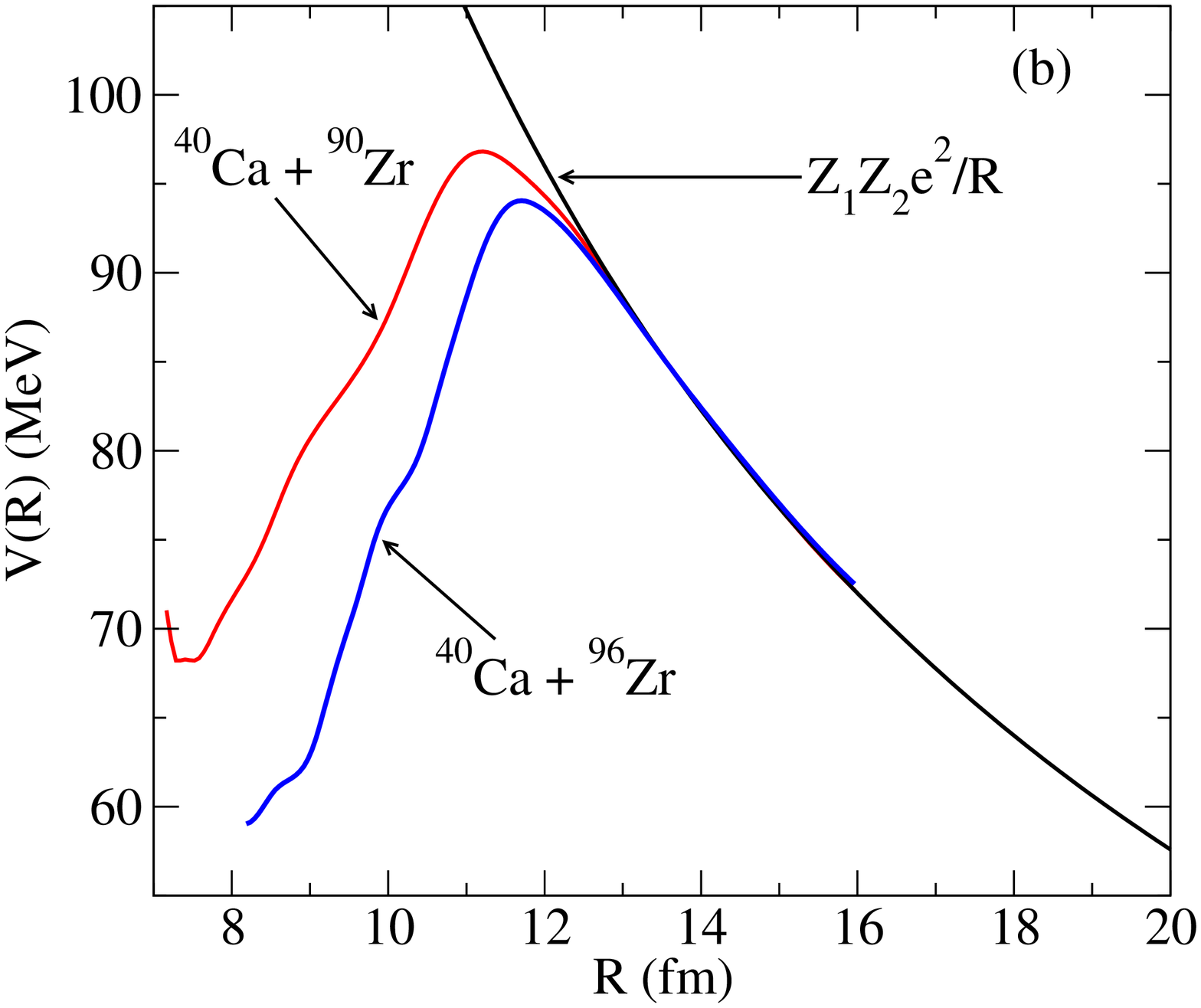}
\caption{\label{fig:vr} (a) Comparison of the DC-TDHF heavy-ion potential for $^{16}$O+$^{16}$O to
the empirical double-folding potential. (b) DC-TDHF heavy-ion potentials for $^{40}$Ca+$^{90,96}$Zr
and comparison to the point Coulomb potential.}
\end{center}
\end{figure*}

In order to test the validity of the initial TDHF setup and the use of the Skyrme interaction without
the c.m. correction term used in the parametrization we utilize the density constrained TDHF method~\cite{CR85,US85}.
In this method we constrain the instantaneous TDHF density and minimize the energy. This in effect
corresponds to the extraction of the internal excitation energy from the evolving system thus tracing
the dynamical trajectory on the multi-dimensional static energy surface of the composite nuclear system.
In Ref.~\cite{UO06b} we have shown that the ion-ion potential barrier could simply be calculated by
subtracting the binding energies of the two nuclei (calculated without the c.m. correction term)
as $V(R)=E_{DC}-E_{A_1}-E_{A_2}$. This in principle tests both the accuracy of the entrance channel
evolution and the energy calculation since for relatively large distances the result must agree with
the point Coulomb potential (for spherical systems). In Fig.~\ref{fig:vr}a we show the results calculated
for the head-on (zero impact parameter) collision of the $^{16}$O+$^{16}$O system at $E_{\mathrm{c.m.}}=34$~MeV.
Also shown is the double folding result for the same system, using densities obtained from
electron-scattering and the M3Y effective NN interaction. In Fig.~\ref{fig:vr}b we show the
barriers calculated for $^{40}$Ca+$^{90,96}$Zr systems. Comparison with the point Coulomb expression
as shown in Fig.~\ref{fig:vr}b is excellent with differences on the order of $50$~keV, until nuclear
effects come into play. Thus, we can conclude that not including the
c.m. correction energy in TDHF calculations does not seem to alter the results when energy differences
are considered. Furthermore, we have also performed the same calculations by changing the
collision energy with no appreciable change in results indicating that the entrance channel TDHF
dynamics is not prone to uncertainties present in the long-time evolution case.

\section{Conclusions}
TDHF calculations in general do not use the center-of-mass correction terms present in most
parametrizations of the Skyrme interaction. In this paper, we discuss this issue along with the problem
of cross-channel coupling. We also provide estimates and numerical tests
to understand the impact of these assumptions. This work, together with the recent investigations
provided in Ref.~\cite{GM08}, where the authors discuss the conservation of angular momentum
in TDHF theory, find that the entrance channel dynamics
of a heavy-ion collision is expected to be well described by the TDHF time evolution.
Also, when energy differences are taken (e.g. in the calculation of heavy-ion potentials)
the omission of the center-of-mass correction terms
do not seem to alter the results. Uncertainties are expected for long-time evolution
resulting in well separated final fragments.

\section*{Acknowledgments}
This work has been supported by the U.S. Department of Energy under grant No.
DE-FG02-96ER40963 with Vanderbilt University

\end{document}